\documentclass[
    ,final            
  ]
  {aipproc}

\layoutstyle{6x9}

\newcommand{\lesssim}{\mathrel{\mathpalette\vereq<}}
\newcommand{\gtrsim}{\mathrel{\mathpalette\vereq>}}
\def\vereq#1#2{\lower3pt\vbox{\baselineskip1.5pt \lineskip1.5pt
  \ialign{$#1\hfill##\hfil$\crcr#2\crcr\sim\crcr}}}

\begin{document}

\title{R-matrix Methods with an application to
${}^{12}{\rm C}(\alpha,\gamma){}^{16}{\rm O}$}


\author{Carl R. Brune}{
  address={Department of Physics and Astronomy, Ohio University, Athens,
  Ohio 45701 (USA)}
}

\begin{abstract}
We review some aspects of R-matrix theory and its application to the
semi-empirical analysis of nuclear reactions.
Important applications for nuclear astrophysics and recent results
for the ${}^{12}{\rm C}(\alpha,\gamma){}^{16}{\rm O}$ reaction are emphasized.
\end{abstract}

\maketitle

\section{Introduction}

The R-matrix theory of reactions was developed by Eugene Wigner in the
1940s~\cite{Wig46a,Wig46b,Wig47} in order to describe resonances in
nuclear reactions.
This approach has proven over
the course of time to be very useful in nuclear and
atomic physics, both for the fitting of experimental data
and as a tool for theoretical calculations.
This paper reviews the use of R-matrix methods for the semi-empirical
analysis of nuclear physics data with an emphasis on situations
that are important for nuclear astrophysics.

The R-matrix theory of nuclear reactions is described in detail
in the review article of Lane and Thomas~\cite{Lan58}.
This method of analysis exactly
incorporates quantum mechanical symmetries and conservation laws
(including unitarity) while at
the same time retaining sufficient empirical flexibility to accurately
describe experimental data.
The long-ranged Coulomb potential is treated exactly.
Modern computational capabilities permit the fitting
of much larger data sets than was previously possible.

Many advances in R-matrix techniques have been motivated by the need
to extrapolate the ${}^{12}{\rm C}(\alpha,\gamma){}^{16}{\rm O}$
cross section to astrophysical energies.
Some recent results on this reaction will be reviewed at the end of
this paper.

\section{R-Matrix Techniques}

Several aspects of R-matrix methods will be reviewed, with an emphasis
on astrophysics and pedagogical aspects.
The notation and mathematical formulas are given by
Lane and Thomas~\cite{Lan58} and only a few essential pieces will be
mentioned here. An R-matrix parameterization is defined by a set of levels
(or equivalently resonances) with well-defined spins and parities.
Each level is described by an energy (related to the excitation energy)
and several reduced width amplitudes.
In general, there is one reduced width amplitude for each channel.
Channels are defined to represent different combinations of nucleons
and/or different combinations of angular momentum coupling.
Finally, for each channel a channel radius and boundary condition
parameter must be specified.
Each level in the R-matrix parameterization is related to a basis
function in the underlying theory.
The main approximations employed are typically truncations of the
numbers of levels and channels considered.
From the R-matrix parameters, the scattering matrix and cross sections can be
computed. Examples of specific computer codes used for fitting and
calculations are described in Refs.~\cite{Lar05,deB08}.

\subsection{When to use R-matrix methods?}

R-matrix methods are most useful in situations where the properties of
individual resonances need to be taken into consideration, as opposed to
cases where other reaction mechanisms dominate or where
resonances can be treated on a statistical basis.
For astrophysical applications, this generally means situations
with light nuclei ($A\lesssim 25$) and low level densities,
or with neutron-induced
reactions at low energies, where individual resonances are also resolved.
This method is particularly useful when more than one resonance of
the same spin and parity are present and the interference effects
between the two states needs to be taken into account.

Since the R-matrix method relies on a semi-empirical parameterization
to describe data, it is necessary that the number of free parameters
be limited in some way and also be well-constrained by data.
One factor which tends to help in this regard is to restrict attention
to low energies -- typically also the astrophysically-interesting choice.
By doing so, the angular momentum barrier limits the number of partial
waves and consequently the number of parameters.
Treating nuclei with low intrinsic spins is also advantageous, as this
limits the complexity of angular momentum coupling and thus also the
number of parameters.
The ${}^{12}{\rm C}(\alpha,\gamma){}^{16}{\rm O}$ reaction
is thus an ideal situation.

The R-matrix approach is also very useful for combining pieces of information
from various reaction channels and other sources.
These sources could include level compilations, lifetime measurements,
transfer reactions, and beta decay.

The use of  R-matrix analyses is driven by applications.
A common use is the extraction of nuclear spectroscopic
quantities (i.e. spins, parities, excitation energies, and partial widths)
from experimental cross section measurements.
Another is the generation of ``best fit'' parameterizations of data for use
in applications.
R-matrix methods are also used for the extrapolation of observables into
unknown (unmeasured) regions of parameter space.
A very common astrophysical use is for the extrapolation of the
cross section or S-factor of a charged-particle reaction to low energies,
as is required for the calculation of a reaction rate in stellar burning.

\subsection{Definition of Resonance Energies}

The level energies in the standard R-matrix approach~\cite{Lan58} are
significantly different from the excitation energies (or resonance energies)
usually employed in nuclear physics.
A different definition of parameters has been developed which uses
the more familiar definition of resonance energy~\cite{Ang00,Bru02}.
I have discussed this method in detail at a JINA workshop~\cite{Bru04}.
This new definition also removes the need for boundary condition
constants which have been a source of confusion in the past.

Even when talking about resonance energies, there are several
different definitions of the resonance energy in use, for example
the point at which the phase shift crosses $\pi/2$, the peak
of a reaction cross section, the R-matrix based definition
discussed in the preceding paragraph, or the complex poles of
the scattering matrix~\cite{Hal87}.
All of these agree for narrow resonances but may diverge as
the resonance width becomes larger.
The important point is to be aware that these different definition
exist and that some differences may just arise from the
definitions used.

\subsection{The Channel Radius}

An important parameter in R-matrix theory is the channel radius.
Nearly all implementations assume that the nuclear force vanishes
for radii larger than this radius.
Nearly all semi-empirical R-matrix fits or parameterizations use
radii which are rather ``small'' -- i.e., such that the nuclear force
has not completely vanished beyond the radius.
For example, most work on fitting the
${}^{12}{\rm C}(\alpha,\gamma){}^{16}{\rm O}$ reaction has used
channel radii between 5 and 7~fm, while theoretical calculations
by Descouvemont and Baye~\cite{Des84} show that a larger radius is
required for the nuclear force to be negligible.

It is not necessarily incorrect to use a ``small'' channel radius
(as defined above) in a semi-empirical calculation.
In fact this approach avoids several problems which arise when a
larger radius is used: the energies of background level(s) are lowered,
a large hard-sphere phase shift must be compensated for by background level(s),
and isospin violation is artificially magnified.
When ``small'' channel radii are used, one must be careful not to
take the parameters to literally (e.g., equating values of reduced widths to
amplitudes of wave functions at the channel radius) for this reason;
quantities related to the asymptotic properties of the wavefunction,
such as observed widths or asymptotic normalization constants,
are not affected by this consideration.

Several studies have found that their conclusions
(e.g., extrapolated S-factors)
are insensitive to the assumed channel radius for a reasonable range
of values (see, e.g., Ref.~\cite{Azu94}).
On the other had, a strong sensitivity of one's results to the channel
radius may suggest that the number of levels in the parameterization is
not sufficient.
Interestingly, the ${}^{12}{\rm C}(\alpha,\alpha)$ experiment of
Tischhauser {\em et al.}~\cite{Tis02} found that a rather small
channel radius of $5.4^{+0.16}_{-0.27}$ gave the best fit to their data.

The question of the channel radius remains open for further study.
One approach would be to use the unified R-matrix plus potential
model of Johnson~\cite{Joh73}. This method allows for a nuclear force
(potential) beyond the channel radius. The penetration and shift factors
are modified from their usual values in this framework.

\subsection{$\beta$ Decay}

R-matrix can be applied to $\beta$-delayed particle energy spectra.
See Azuma {\em et al.}~\cite{Azu94} for a measurement and analysis
of the $\beta$-delayed $\alpha$ spectrum following the decay
of ${}^{16}{\rm N}$. This analysis places useful constraints on
the ${}^{12}{\rm C}(\alpha,\gamma){}^{16}{\rm O}$ reaction.

The formula for analysis have been given by Barker and Warburton~\cite{Bar88}.
However, no derivation has been given for their result
and it is not clear what assumptions are necessary.
Also, their work was limited to the study of allowed transitions.

\subsection{Transfer Reactions}

Transfer reactions are a very useful tool for studying states of astrophysical
interest. In addition to the excitation energy, spin, and parity,
transfer reactions can determine the spectroscopic factor or reduced
particle width of a state.
The typical manner of analysis of transfer reaction data is to use a
distorted-wave born approximation (DWBA) calculation for the cross section.
The normalization of the calculation to the data determines the
spectroscopic factor.
The spectroscopic factor, along with the bound-state wavefunction used in
the DWBA calculation, can then be used to determine the reduced width
for an R-matrix calculation.

For the case of bound states, the results of a transfer reaction
measurements can be described by an asymptotic normalization constant (ANC).
This technique, and the relationship between the ANC and the R-matrix
reduced width, is discussed in detail
by Mukhamedzhanov and Tribble~\cite{Muk99}.
Both a transfer reaction and an R-matrix description are primarily
concerned with the long-ranged parts of wavefunctions.
The primary advantage of using the ANC approach is that the
spectroscopic factor and its associated uncertainties
due to the poorly-known short-range part of the bound-state
wavefunction do not enter.

For unbound states, a similar method has been described
by Iliadis~\cite{Ili97} which leads to the partial
width (or R-matrix reduced width) of interest.
In this approach, the partial width is related to the product of
a spectroscopic factor and a single-particle width.
Provided that exactly the same potential is used for the
DWBA calculation and the single-particle width calculation,
the uncertainty in the spectroscopic factor due to the short-ranged
part of the wavefunction does not contribute to the error in
the final partial width.

Attention must be paid to several details when relating transfer
reaction results to R-matrix parameters.
One such detail is the difference between formal and partial widths (or reduced
widths); see Iliadis~\cite{Ili97}.
Another is that the spectroscopic factor
depends on whether the single-particle wavefunctions are
normalized to unity inside the channel radius, or over all space
(see Ref.~\cite{Bar97} and the distinction between ``spectroscopic
factor'' and ``R-matrix spectroscopic factor'' raised in Ref.~\cite{Moh97}).
DWBA calculations typically normalize the single-particle wavefunction
to unity over all space.
An alternative method for converting spectroscopic factors determined with
transfer reactions to reduced widths has been given by
Becchetti {\em et al.}~\cite{Bec78}.
It should be noted that these equations are only valid if the nuclear part
of the bound-state potential used in the DWBA calculation vanishes beyond
the channel radius or if modified penetration factors are utilized
(as per the unified R-matrix plus potential
model of Johnson~\cite{Joh73} discussed above).
Neither of these conditions are typically employed.

\subsection{Radiative Capture}

The inclusion of photon channels is discussed in the work of
Lane and Thomas~\cite{Lan58}.
An important extension has been provided by Holt~\cite{Hol78}
and Barker and Kajino~\cite{Bar91}.
Their treatment includes contributions to the radiative capture matrix
elements from radii beyond the channel radius, such as occurs
with the direct capture mechanism.
These contributions are particularly important for situations where
the capture cross section is non resonant.

\section{The ${}^{12}{\rm C}(\alpha,\gamma){}^{16}{\rm O}$ Reaction}

The $^{12}{\rm C}(\alpha,\gamma)^{16}{\rm O}$ reaction plays
a very important role in stellar evolution.
The rate of this reaction determines the $^{12}{\rm C}/^{16}{\rm O}$ ratio
produced by helium burning, and consequently has very significant
effects on the subsequent structure and nucleosynthesis, as
well as the final outcome of the evolution~\cite{Wea93,Has95}.
Calculation of the reaction rate in helium-burning conditions requires
the cross section be known for energies of
$\approx 0.3$~MeV.
However, the cross section has been measured in the laboratory
only for energies $\gtrsim 1$~MeV.
Some aspects of the ${}^{12}{\rm C}(\alpha,\gamma){}^{16}{\rm O}$
reaction are reviewed here.

\subsection{Separating the $E1$ and $E2$ Ground State Components}

The $E1$ and $E2$ radiative capture processes to the $0^+$
${}^{16}{\rm O}$ ground state proceed via $1^-$ and $2^+$ states, respectively.
These processes may produce interference effects in the angular distribution.
The formula for the angular distribution of the
${}^{12}{\rm C}(\alpha,\gamma_0)^{16}{\rm O}$ reaction has been given
by Dyer and Barnes~\cite{Dye74}:
\begin{eqnarray}
{{\rm d}\sigma\over{\rm d}\Omega}&=&{1\over 4\pi}\biggl[ \sigma_{E1}(1-P_2)
+\sigma_{E2}(1+{5\over 7}P_2-{12\over 7}P_4) \nonumber \\
&&+6\cos\phi\sqrt{\sigma_{E1}\sigma_{E2}\over 5}(P_1-P_3)\biggr],
\label{eq:dsdo}
\end{eqnarray}
where the angular dependence is contained in the
Legendre polynomials $P_l\equiv P_l(\cos\theta)$
and $\phi$ is a phase parameter.
The term $\propto\sqrt{\sigma_{E1}\sigma_{E2}}$ results from $E1$-$E2$
interference, and gives rise to an asymmetry around $90^\circ$
in the differential cross section.

Equation~\ref{eq:dsdo} has been used by several experiments;
see for example the recent work of Assun\c{c}\~{a}o {\em et al.}~\cite{Ass06}.
The phase parameter $\phi$ is determined by the difference between
the $L=1$ and $L=2$ nuclear and Coulomb phase shifts~\cite{Bru01}.
This quantity can be fixed much more accurately by
${}^{12}{\rm C}({\alpha,\alpha)}$ measurements, particularly now that
new high-precision data are available~\cite{Tis02,Tis09}.
These data will permit a more accurate separation of the $E1$ and $E2$
components, or, in the case of a simultaneous fit, a better constraint on the
fitted parameters.

Additional work on the $E1$ and $E2$ ground-state components would
help to constrain the extrapolation of the 
${}^{12}{\rm C}(\alpha,\gamma){}^{16}{\rm O}$ S-factor
to astrophysical energies.
There is very little data above $E_{c.m.}=3$~MeV; measurements here
would help to constrain the impact of higher-lying energy levels.
Another interested possibility is the a measurement of the integrated
angular distribution of the narrow resonance at $E_{c.m.}=2.68$~MeV.
As shown by Brune~\cite{Bru01}, such a measurement could shed light
on how the 2.68-MeV resonance interferes with other $E2$ processes.

\subsection{Recent Progress}

There has been some significant progress on the
${}^{12}{\rm C}(\alpha,\gamma){}^{16}{\rm O}$ reaction in recent years.
The total cross section has been determined by Sch{\" u}rmann
{\em et al.}~\cite{Sch05} using a recoil separator.
These data have very small statistical errors and place a
constraint on the sum of capture components.
New angular distributions of the ground-state cross section
have been reported by Assun\c{c}\~{a}o {\em et al.}~\cite{Ass06} which appear to have
optimized the approach of using an intense $\alpha$ beam and Germanium
$\gamma$-ray detectors.
The first measurements of the capture into the 6.05-MeV exited state
were performed by Matei {\em et a.}~\cite{Mat06}.
This work suggests that the extrapolation of this cascade cross section
may be higher than previously thought.
A new determination of the $\beta$-delayed $\alpha$ spectrum from
${}^{16}{\rm N}$ as been made by Tang {\em et al.}~\cite{Tan07}.
This measurement essentially confirms the previous result~\cite{Azu94}
and consequently also the constraint on the $E1$ ground-state cross section.
Finally, a tight limit on the $\gamma$-ray branching ratio
of the 7.12-MeV state of ${}^{16}{\rm O}$ into the 6.92-MeV state
has been recently found~\cite{Mat08}.
This branching ratio impacts the cascade cross section into the
6.92-MeV state, which in turn impacts the reduced $\alpha$
width determined for this state.
It is anticipated that this progress will continue into the future
and that further improvements in our understanding of the
rate of ${}^{12}{\rm C}(\alpha,\gamma){}^{16}{\rm O}$ in red
giant stars will come to pass.


\begin{theacknowledgments}
It is a pleasure to thank Daniel B. Sayre for useful comments
on the manuscript.
This work is supported in part by the U.S. Department of Energy
under grant no. DE-FG02-88ER40387.
\end{theacknowledgments}

\bibliographystyle{aipproc}   

\end{document}